# Quantification of lung function on CT images based on pulmonary radiomic filtering

Zhenyu Yang[1,2], Kyle J Lafata[1,3,4], Xinru Chen[2], James Bowsher[2], Yushi Chang[1], Chunhao Wang[1], and Fang-Fang Yin[1,2*]

[1]Deparment of Radiation Oncology, Duke University, Durham, NC, 27710, United States
[2]Medical Physics Graduate Program, Duke Kunshan University, Kunshan, Jiangsu, 215316, China
[3]Department of Radiology, Duke University, Durham, NC, 27710, United States
[4]Department of Electrical and Computer Engineering, Duke University, Durham, NC, 27710, United States

**Short Running Title: Pulmonary radiomic filtering**

*Corresponding author:
Fang-Fang Yin, Ph.D.
Department of Radiation Oncology
Duke University Medical Center
Durham, NC, 27710, United States
E-mail: fangfang.yin@duke.edu

# Abstract

**Purpose**: To develop a radiomics filtering technique for characterizing spatial-encoded regional pulmonary ventilation information on lung CT.

**Methods**: The lung volume was segmented on 46 CT images, and a 3D sliding window kernel was implemented across the lung volume to capture the spatial-encoded image information. Fifty-three radiomic features were extracted within the kernel, resulting in a $4^{th}$ order tensor object. As such, each voxel coordinate of the original lung was represented as a 53-dimensional feature vector, such that radiomic features could be viewed as feature maps within the lungs. To test the technique as a potential pulmonary ventilation biomarker, the radiomic feature maps were compared to paired functional images (Galligas PET or DTPA-SPECT) based on Spearman correlation ($\rho$) analysis.

**Results**: The radiomic feature map *GLRLM-based Run-Length Non-Uniformity* and *GLCOM-based Sum Average* are found to be highly correlated with the functional imaging. The achieved $\rho$ (median [range]) for the two features are 0.46 [0.05, 0.67] and 0.45 [0.21, 0.65] across 46 patients and 2 functional imaging modalities, respectively.

**Conclusions**: The results provide evidence that local regions of sparsely encoded heterogeneous lung parenchyma on CT are associated with diminished radiotracer uptake and measured lung ventilation defects on PET/SPECT imaging. These findings demonstrate the potential of radiomics to serve as a complementary tool to the current lung quantification techniques and provide hypothesis-generating data for future studies.

# Introduction

Lung disease is a leading cause of global mortality [1, 2], and assessment of pulmonary function is essential in patients with lung disease [3, 4]. Ventilation denotes the exchange of blood and gas in the lungs and is a common surrogate of lung function [3]. The identification of the ventilation defects helps screen lung diseases [4], evaluate radiation therapy treatment plans [5], and minimize treatment-induced parenchymal injury [6]. Over the past few decades, nuclear medicine-based ventilation scintigraphy has served as the clinical gold standard for tomographic pulmonary function imaging and assessment [7]. Typically, nuclear medicine-based scintigraphy is costly, and prophylactic applications are limited due to commonly observed adverse side effects [8]. Such limitations, along with the growing socioeconomic burdens linked to lung disease, have motivated the development of new quantitative imaging techniques in recent years [9].

Advances in new techniques to characterize high-dimensional feature spaces [10] and quantitative applications of CT imaging – which is currently the standard of care for lung imaging - suggest that lung disease and pulmonary function can be treated as a regionally heterogeneous system [9, 11, 12]. Earlier efforts in quantitative pulmonary function assessment include Hounsfield Unit (HU)-based thresholding, but the reported performance was limited [12]. Computed tomography ventilation imaging (CTVI) is an emerging pulmonary function quantification method using respiratory-gated four-dimensional CT (4DCT). In some reported studies, deformable image registration (DIR) and motion modeling analysis [13-15] were used in CTVI to characterize breathing-induced changes in lung density and air volume [3, 9, 14]. Such procedure not only relies on the acquisition of 4DCT, but also its results can be sensitive to the choice of DIR algorithm and motion model [5, 16-18].

Radiomics is a popular medical imaging quantitative framework, which employs feature engineering techniques to convert standard-of-care medical images into mineable data [19]. Previous radiomic-based pulmonary studies reported that the intensity and texture features of the lung are associated with global pulmonary function measurements, such as forced expiratory volume in one second ($FEV_1$) and diffusing capacity of the lung for carbon monoxide (DLCO) [11]. However, radiomic analysis of the entire lung volume as a region-of-interest (ROI) is insufficient to characterize spatial-encoded pathological and physiological characteristics of the lungs. Spatial information is critical for the clinical interpretation of lung function, e.g., in the preoperative assessment of lung resection candidates with impaired lung function [20]. The sub-regional radiomic analysis, known as habitat analysis, combined with machine learning and deep learning, has been reported to identify the lung function [21-23]. However, the direct analysis of true spatial-encoded radiomic features with the regional pulmonary ventilation is needed.

In this work, a radiomics filtering technique is developed to capture spatial-encoded image pathological and physiological features associated with underlying ventilation changes from lung CT. Motivated by habitat radiomics, the developed radiomic filtering technique extends the radiomic extraction to a voxel basis. To test the radiomic filtering of the lungs, we applied our technique to a publicly available CT dataset, where nuclear medicine-based ventilation images were also available as reference measurements. The developed technique may serve as a complementary tool to current pulmonary quantification techniques and provides hypothesis-generating data for future studies in quantitative pulmonary imaging.

# Methods

The following methods were carried out under relevant regulations. Retrospective data analysis was completed with approval from the Duke University Health System Institutional Review Board. All data is de-identified and is publicly available as part of the VAMPIRE Challenge dataset [3].

### A.     Imaging data

A public non-small cell lung cancer (NSCLC) patient dataset from the VAMPIRE Challenge [3] was used as the basis of this work. Specifically, the VAMPIRE dataset includes paired image acquisitions of CT and reference ventilation images (RefVI) crossing ventilation imaging modalities and clinical settings [3]. Forty-six lung cancer patients were studied in this work, including 25 patients with Galligas 4DPET/CT imaging and 21 patients with DTPA-SPECT/CT imaging. 4DCT images were binned into (1) 5 respiratory phases with a resolution of 1.07 (mm) × 1.07 (mm) × 5.00 (mm, slice thickness) for the 25 4DPET/CT patients, and (2) 10 phases with a resolution of 0.97 (mm) × 0.97 (mm) × 2.0, 2.5 or 3 (mm, slice thickness) for the 21 DTPA-SPECT/CT patients. All CT images were acquired under free-breathing conditions. Average CT volumes were derived for all 46 patients and finally adopted for our radiomic analysis. All RefVIs (PET and SPECT) were obtained after CT acquisition and were co-registered to the corresponding average CT image. For each patient, lung segmentation masks were included in the dataset for both the average CT image and the RefVI image.

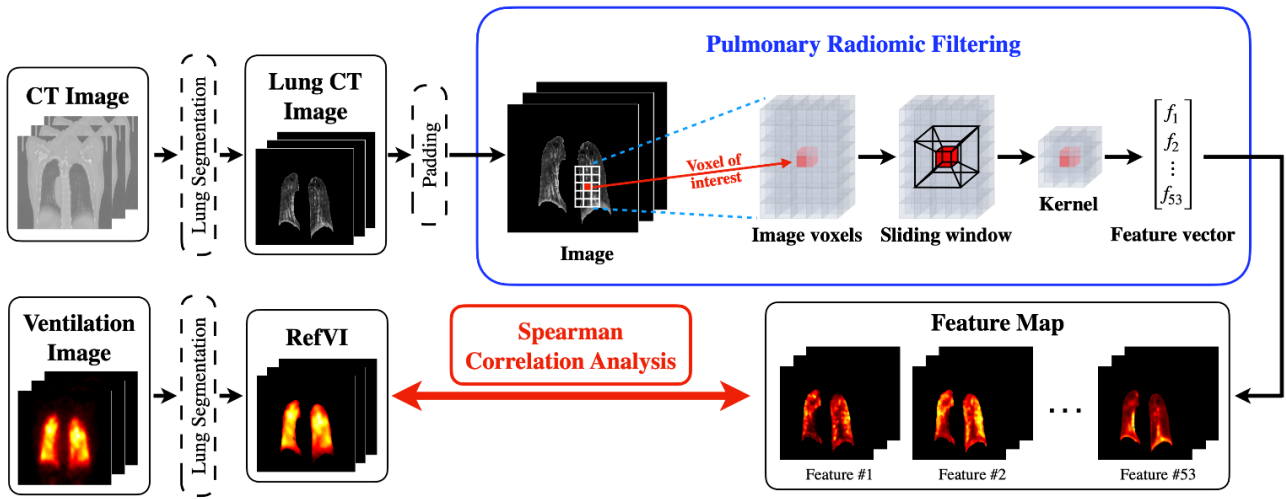

Figure 1. The overall design of this work.

## B. Radiomic filtering

The overall research design and methodological workflow, including radiomic filtering, feature map analysis, and functional imaging comparison, is summarized in Figure 1. The lung volume was first obtained from the average CT image and the segmentation mask for each patient. A 3D sliding window kernel was implemented to capture the regional radiomic features throughout the entire lung volume. Accordingly, each voxel coordinate in the original lung was represented as an $n$-dimensional feature vector, and the feature space can be represented as an $n$-dimensional 4$^{th}$ order tensor, i.e., a set of shift-invariant 3D *feature maps*. Due to this shift-invariance property, each feature map can be viewed in the same reference frame as the original CT image. Formally, given $n$ radiomic features and a CT image with a dimension of $x \times y \times z$, the feature maps $\mathcal{F}$ can be mathematically represented as a multi-dimensional tensor:

$$\mathcal{F} = (f_{i,j,k,l}) \in \mathbb{R}^{x \times y \times z \times n} \tag{1}$$

where the $(i,j,k,l)^{th}$ component denotes the measured value of the $l^{th}$ radiomic feature at the $(i,j,k)^{th}$ tomographic coordinate.

In this work, a total of $n = 53$ radiomic features were included. These features were chosen to collectively capture the local texture intensity and texture characteristics within the lungs [19]. Table I summarizes all 53 features, which can be grouped into three types based on different joint-probability functions:
1) Gray level co-occurrence matrix (GLCOM)-based features: measure the distribution of pair-wise gray level combinations in the image.
2) Gray level run-length matrix (GLRLM)-based features: measure the distribution of consecutive intensity values of the same gray level in a specified direction in the image.
3) Grey level size zone matrix (GLSZM)-based features: measure the distribution of pixel patches of a similar gray level within the image.

| | | | | | |
|---|---|---|---|---|---|
| Grey level co-occurrence matrix (GLCOM)-based features | 1 | Auto Correlation | Grey level run-length matrix (GLRLM)-based features | 28 | Run Percentage |
| | 2 | Cluster Prominence | | 29 | Low Gray Level Run Emphasis |
| | 3 | Cluster Shade | | 30 | High Gray Level Run Emphasis |
| | 4 | Cluster Tendency | | 31 | Short Run Low Gray Level Emphasis |
| | 5 | Contrast | | 32 | Short Run High Gray Level Emphasis |
| | 6 | Correlation | | 33 | Long Run Low Gray Level Emphasis |
| | 7 | Differential Entropy | | 34 | Long Run High Gray Level Emphasis |
| | 8 | Dissimilarity | | 35 | Grey Level Variance |
| | 9 | Energy | | 36 | Run Length Variance |
| | 10 | Entropy | | 37 | Run Entropy |
| | 11 | Homogeneity 1 | | 38 | Small Zone Emphasis |
| | 12 | Homogeneity 2 | Grey level size zone matrix (GLSZM)-based features | 39 | Large Zone Emphasis |
| | 13 | Info Measure Correlation 1 | | 40 | Gray Level Non-Uniformity |
| | 14 | Info Measure Correlation 2 | | 41 | Gray Level Non-Uniformity Normalized |
| | 15 | Inverse Difference Moment Normalized | | 42 | Size Zone Non-Uniformity |
| | 16 | Inverse Difference Moment | | 43 | Size Zone Non-Uniformity Normalized |
| | 17 | Inverse Variance | | 44 | Zone Percentage |
| | 18 | Maximum Probability | | 45 | Low Gray Level Size Emphasis |
| | 19 | Sum Average | | 46 | High Gray Level Size Emphasis |
| | 20 | Sum Entropy | | 47 | Small Size Low Gray Level Emphasis |
| | 21 | Sum Variance | | 48 | Small Size High Gray Level Emphasis |
| Grey level run-length matrix (GLRLM)-based features | 22 | Short Run Emphasis | | 49 | Large Size Low Gray Level Emphasis |
| | 23 | Long Run Emphasis | | 50 | Large Size High Gray Level Emphasis |
| | 24 | Gray Level Non-Uniformity | | 51 | Gray Level Variance |
| | 25 | Gray Level Non-Uniformity Normalized | | 52 | Zone Size Variance |
| | 26 | Run-Length Non-Uniformity | | 53 | Zone Size Entropy |
| | 27 | Run-Length Non-Uniformity Normalized | | | |

Table I. Fifty-three radiomic features included in this study.

## C. Calculation

The radiomic filtering was performed with respect to the 4DCT average lung volume. Following the previous lung radiomic studies, the dynamic range of the lung was re-binned to 64 gray levels [11, 24, 25]. The choice of sliding window (kernel) size defines the radiomic filtering resolution. A small kernel may lead to a noisy output feature map, while a large kernel may lose the signal. The typical pulmonary defect volumes vary from $20 \text{ cm}^3$ to $80 \text{ cm}^3$, and the diameter is 20-40 mm [26, 27]. Here, we set a kernel size of $15 \times 15 \times 15 \text{ mm}^3$, which is on the order of a typical pulmonary defect. All 53 features were averaged over 13 directions [24] to approximate rotational invariance [28-31].

All the calculations were performed using an in-house developed radiomics filtering toolbox with MATLAB (MATLAB R2018a; MathWorks, Natick, Mass). The toolbox was fully validated against the image biomarker standardization initiative (IBSI) standardization [32] as well as the digital phantoms [33]. In addition, the toolbox was specifically optimized for this voxel-based, rotationally invariant, 3D calculation. It included (1) fully vectorized calculation procedures to accelerate small-kernel matrix manipulation and (2) parallel computing to accelerate distributed computation. The average calculation time for each patient was about 10 minutes, which is effective considering that the operation was performed on the entire 3D lung volume at native resolution.

## D. Multi-collinearity assessment

To reduce feature redundancy in equation (1), we performed a dimensionality reduction procedure as follows. For each patient, Pearson correlation coefficients ($r$) of each feature pair were calculated, thereby leading to 46 covariance matrices. The mean Pearson $\bar{r}$ covariance matrix was considered as a representation of the multi-collinearity existing in $\mathcal{F}$, and the feature maps with $\bar{r} > 0.95$ were considered as highly correlated [19, 34, 35]. The highly correlated features were classified as a feature subset and treated as a group in the subsequent analysis.

### E. Correlation analysis

The extracted radiomic feature maps were quantitatively compared to the corresponding RefVI. For each feature map/RefVI matched pair, voxel-wise Spearman correlation coefficients ($\rho$) were calculated within the lungs. Prior to the $\rho$ calculation, all feature maps were resampled to match the spatial dimensions of the RefVI via a nearest-neighbor interpolation. To keep consistency with the VAMPIRE study, the RefVI images were smoothed as follows: voxels with ventilation intensity above $\pm 4$ standard deviations of the mean intensity of overall RefVI lung voxels were removed until the threshold converged to within 1% of the last threshold [3, 36].

### F. Comparison study

We hypothesized that the joint measurement of image intensity and texture by radiomics analysis would provide complementary information compared to intensity-only (e.g., thresholding) analysis [25]. To test the hypothesis, we performed the following three intensity-based studies:

1) The Spearman correlation between original lung CT with RefVI.
2) The Spearman correlation between average filtering lung CT with RefVI. The average filter calculates the regional mean within a $15 \times 15 \times 15 \text{ mm}^3$ sliding window kernel for comparison with radiomics filtering.
3) The regional overlap between the pulmonary defects was identified by intensity thresholding with ground truth defects. The CT value of $-950$ HU is considered as an acceptable threshold between emphysema and normal lung [12]. According to the VAMPIRE study, the region with the lowest 25% percent of the total intensity in PET/SPECT is considered as the ground truth pulmonary defects [3]. The Dice coefficient was employed to quantify the regional overlap.

The Spearman correlations from the first and second studies were evaluated against the highest-performing radiomic results. Wilcoxon signed-rank tests were performed on the Spearman coefficient, where p<0.05 was considered statistically significant.

# Results

Figure 2 provides an illustrating example of 53 radiomic filtered feature maps for a single patient. The upper left panel shows the input average CT image and the corresponding RefVI. The RefVI is normalized to [0, 1] and is superimposed over the CT at a common coronal location for better visualization. A major ventilation defect is clearly demonstrated in the left upper lobe (LUL) of RefVI. The remaining panels are the extracted 53 feature maps with feature numbers in the top left corner. Similarly, each feature map is normalized with the same approach as the RefVI to provide a normalized dynamic range. Several radiomic feature maps (e.g., #19 *GLCOM-based Sum Average* and #26 *GLRLM-based Run-Length Non-Uniformity*), exhibit reasonable visual concordance with the RefVI at the defect location. These features are potentially directly proportional to lung ventilation. Meanwhile, some other features present distinguished high values in the defects, including feature #9, #29, #31, and #33; these features may be inversely proportional to the lung ventilation measurement.

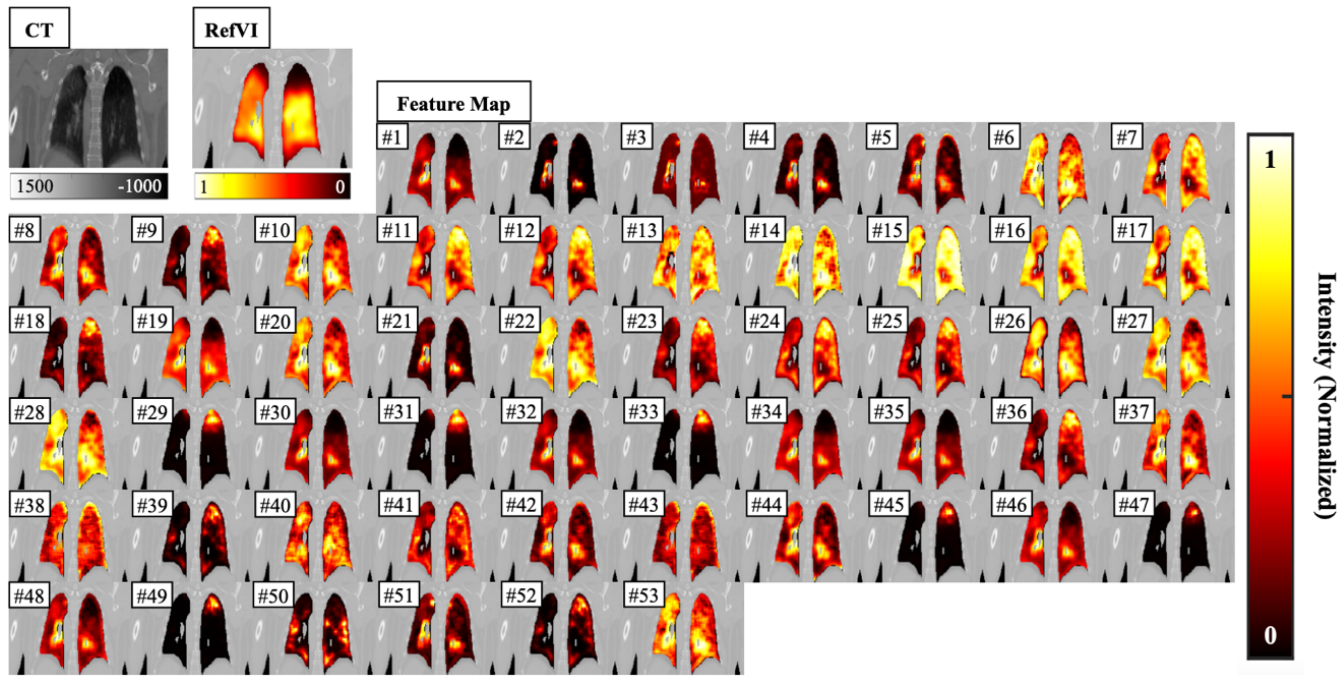

Figure 2. Visual inspection of 53 radiomic feature maps. The upper left panels are the original input averaged CT volume and corresponding RefVI slice.

Table II summarizes the multi-collinearity assessment results. Eighteen features are identified as independent, and the rest 39 features are grouped as 12 subsets. The features included in each subset are highly correlated (mean Pearson $\bar{r} > 0.95$).

|  | Feature # |
|---|---|
| **Feature subset 1** | 19, 52 |
| **Feature subset 2** | 9, 18 |
| **Feature subset 3** | 11, 12, 17, 23, 36 |
| **Feature subset 4** | 15, 16 |
| **Feature subset 5** | 29, 31, 33 |
| **Feature subset 6** | 45, 47 |
| **Feature subset 7** | 38, 43 |
| **Feature subset 8** | 22, 27, 28 |
| **Feature subset 9** | 10, 20, 37 |
| **Feature subset 10** | 8, 44 |
| **Feature subset 11** | 1, 30, 32, 34, 35, 46, 48 |
| **Feature subset 12** | 4, 21 |
| **Independent Features** | 2, 3, 5, 6, 7, 13, 14, 19, 24, 25, 26, 40, 41, 42, 49, 50, 51, 53 |

Table II. Results of multi-collinearity assessment.

Figure 3 shows the Spearman correlation analysis results. Figure 3 (a)-(c) are the $\rho$ distributions evaluated between 18 independent features and corresponding RefVI scans for the full 46 patients, 25 Galligas 4D PET/CT patients, and 21 DTPA-SPECT patients, respectively. Figure 3 (d)-(f) are the $\rho$ distributions of 12 feature subsets for full 46 patients, Galligas 4D PET/CT patients, and DTPA-SPECT patients, respectively. The features are ranked in descending order from left to right according to the median value of $\rho$. The highest correlations are found to be two independent features: (1) feature #26 *GLRLM-based Run-Length Non-Uniformity* and (2) feature #19 *GLCOM-based Sum Average*. As highlighted with the green boxes, these two features show a relatively robust correlation across 46 patients and 2 nuclear medicine-based ventilation imaging modalities. Specifically, the *GLRLM-based Run-Length Non-Uniformity* achieves $\rho$ median [range] of 0.45 [0.21, 0.65] for full 46 patients, 0.45 [0.29, 0.74] for 25 Galligas PET patients, and 0.45 [0.21, 0.65] for 21 DTPA-SPECT patients, respectively. *GLCOM-based Sum Average* reaches 0.46 [0.05, 0.67], 0.44 [0.13, 0.74], 0.44 [0.13, 0.74] for full patients, Galligas PET patients, and DTPA-SPECT patients, respectively.

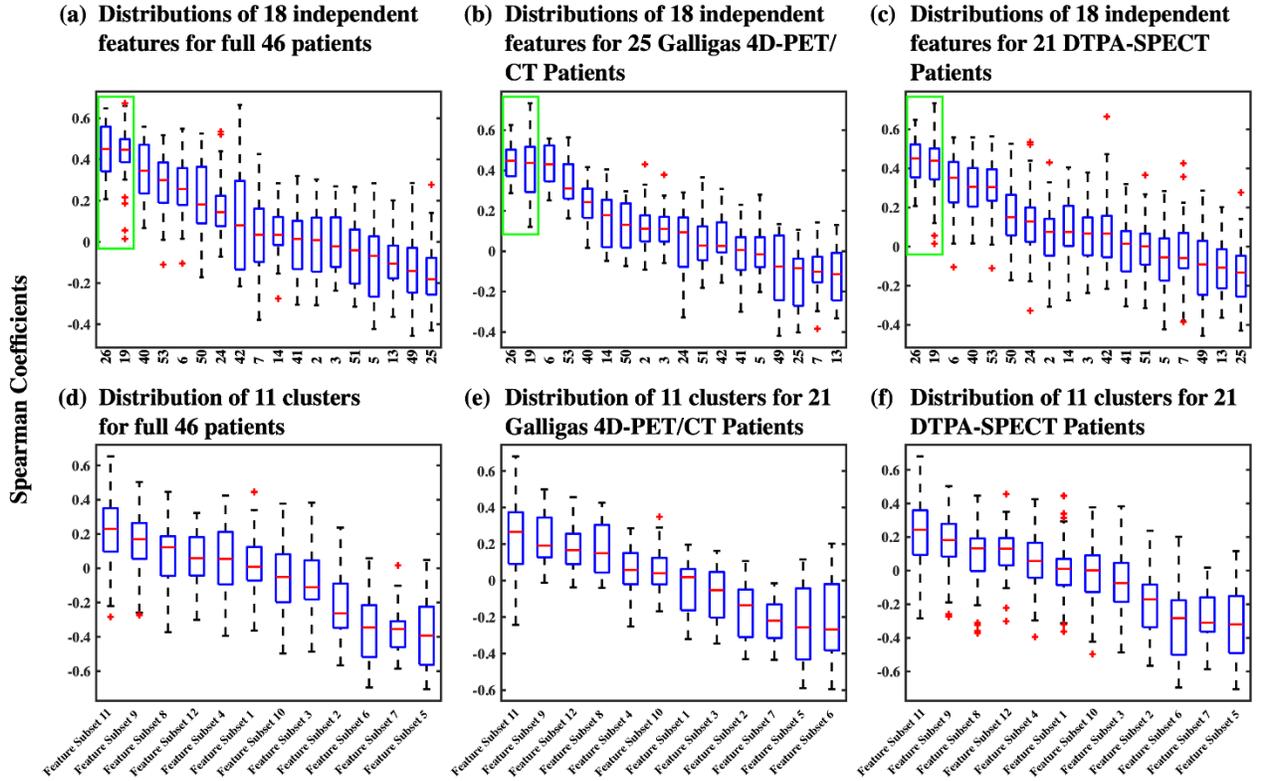

Figure 3. The Spearman correlation analysis results of 18 independent features and 12 feature subsets. In each box, the central mark indicates the median, and the bottom and top edges denote the 25th and 75th percentiles, respectively. The whiskers extend to the most extreme data points not considered as outliers, and the outliers are plotted individually using the "+" symbol. The highest-performing features were highlighted with green boxes.

Table III summarizes the results of the HU-based analysis. The measured association in all three studies is very low. The Spearman coefficients in the first and second studies are significantly lower than the highest-performing radiomic results.

|  | 25 PET/CT patients | 21 DTPA-SPECT patients |
| --- | --- | --- |
| **Study 1 (Spearman)** | 0.22 [0.13, 0.34]* | 0.18 [0.10, 0.30]* |
| **Study 2 (Spearman)** | 0.31 [-0.01, 0.46]* | 0.29 [0.12, 0.51]* |
| **Study 3 (Dice)** | 0.24 [0.11, 0.42] | 0.20 [0.14, 0.38] |

Table III. Results of three HU-based studies (median [range]). "*" indicates a statistically significant difference compared with the highest-performing radiomics results.

## Discussion

The assessment of regional pulmonary function is essential in patients with lung disease. Modern imaging systems – in combination with state-of-the-art computational techniques (e.g., radiomics) – enable new approaches to this problem. In this work, we hypothesize that spatially encoded radiomic textures are inherently associated with local changes in pulmonary function. To test the hypothesis, we developed a radiomic filtering workflow to characterize the association between regionally heterogeneous lung CT information and pulmonary ventilation measured by RefVI (PET and SPECT). Based on the VAMPIRE dataset, the radiomic filtering feature maps with the highest correlation to RefVI are *GLRLM-based Run-Length Non-Uniformity* (#26) and *GLCOM-based Sum Average* (#19). *GLRLM-based Run-Length Non-Uniformity* assesses the distribution of runs over the run lengths. A lower value indicates high homogeneity in run-length measured at the source pixel location. *GLCOM-based Sum Average* is the sum of the gray level with voxel-pairs distributions, which measures the relationship between sparse and dense areas in an image [37]. The HU value in a certain lung region can be considered as a linear combination of a water-like material and air-like material [38-40]. Hence, the *GLCOM-based Sum Average* can be taken as a measurement of the local air content change [41, 42]. It is generally known that the reduction of pulmonary function is closely associated with diffuse alveolar epithelium destruction, capillary damage/bleeding, hyaline membrane formation, alveolar septal fibrous proliferation, and pulmonary consolidation [43-45]. The reduction in lung function can be assumed as the healthy pulmonary tissue is replaced by an increase of air within the lungs [25]. As such, the reduction in lung function leads to regional heterogeneity and a decrease in CT attenuation coefficient, which could be potentially reflected as lower *GLRLM-based Run-Length Non-Uniformity* and lower *GLCOM-based Sum Average*. Collectively, these two features suggest that local regions of sparsely encoded heterogeneous lung parenchyma on CT are associated with diminished radiotracer uptake and measured lung ventilation defects on PET/SPECT imaging. This finding is consistent with previous whole lung-based analyses of the emphysema [46], asthma [47], lung inflammation [48], COPD [49], and fibrosis [50].

The VAMPIRE Challenge benchmarked 37 independent CTVI algorithms using the VAMPIRE dataset based on the Spearman coefficient [3]. Since this work employed the same dataset as well as the statistical analysis method, the Spearman in the VAMPIRE Challenge can be considered a reasonable benchmark for new hypothesis-generating data. The benchmark results of three best performing CTVI algorithms in the VAMPIRE Challenge achieves $\rho$ (median [range]) of 0.49 [0.27, 0.73], 0.38 [-0.10, 0.65], and 0.37 [- 0.20, 0.60] for full 46 patients, Galligas PET patients, and DTPA-SPECT patients, respectively. In our study, the top correlated feature *GLRLM-based Run-Length Non-Uniformity* (#26) and *GLCOM-based Sum Average* (#19) obtained comparable results. Compared to intensity-only analysis, the radiomic filtering outperforms the raw HU value and CT intensity average filtering. HU thresholding is also limited in capturing pulmonary defects. We note that other advanced intensity analyses have been reported to improve lung function identification performance [51]. Our comparison study suggests that incorporating CT texture features may provide complementary information related to lung function. The developed radiomic filtering technique potentially extends the conventional radiomic approach without losing spatial-encoded information, and can be easily integrated as input into the other lung function quantification workflows, such as CTVI, machine learning, and deep learning.

Although this research has demonstrated promising results and has provided hypothesis-generating data, our study also has several limitations. First, the radiomic filtering technique may also capture the non-parenchymal voxel intensities (e.g., blood vessels and artifacts). The VAMPIRE dataset has excluded the major blood vessels from the lung contours. The pulmonary vessels with a cross-sectional diameter of < 5 mm are also shown to be related to emphysema and pulmonary hypertension [52, 53]. We hypothesized that the small blood vessels could also be regarded as a regional heterogeneous system [54], closely related to lung ventilation and airflow, which can be captured by radiomics. We note that such vessel segmentation masks, as provided by the VAMPIRE, largely preclude the drawbacks of DIR-based CTVI (e.g., extended processing time). The incorporation of DIR into image fusion software and radiotherapy treatment planning systems also reduces the need for manual expertise operation. It is worth exploring the radiomic filtering on pulmonary vasculature images, which could potentially expand the utility and reach of the technique. Second, due to the limitation of the current VAMPIRE dataset, the proposed radiomic filtering technique was designed and implemented in 4DCT average volume. The averaged 4DCT volume is not a smooth encoding of respiratory motion, it is a numerical average of phase binned segments sorted according to respiratory trace information obtained over a long scan time. It has been reported that 4DCT average lung volume is highly consistent with free-breathing CT in radiotherapy contouring, planning, and dose calculation for lung cancer patients [55]. However, the performance of the proposed radiomic filtering technique on free-breathing CT remains unknown. The sorting artifacts caused by lung motion may lead to locally unresolvable uncertainties in the radiomics filtering feature maps. Further characterization on the impact of 4DCT averaged images, phase images, free-breathing images, and breath-hold images on voxel-wise feature quantification are warranted, particularly minimum resolvable texture. Finally, the variability of possible CT acquisition settings can also introduce uncertainties for the radiomic analysis. Typically, the radiomic analysis is sensitive to institution-specific differences in kV tube settings (kVp/mAs), scan collimation/slice thickness, consistency of HU calibration, choice of standard or iterative reconstruction, and the prevalence of 4DCT motion artifacts (e.g., different types of phase/amplitude binning, and selection of scan pitch) [56-60]. In this study, the fully curated VAMPIRE dataset has relatively uniform acquisition parameters, signal-to-noise ratio, etc. The filtering process focuses on the relative value of the feature in a regional kernel, rather than its absolute values, which may be less sensitive than the traditional radiomic approach. However, a more comprehensive generalization and robustness analysis of our technique should be further characterized as part of future work.

## Conclusion

In this work, we presented a computational radiomic filtering technique to quantify local changes in pulmonary function on CT images. Our results provide evidence that local regions of sparsely encoded heterogeneous lung parenchyma on CT are associated with diminished radiotracer uptake and measured lung ventilation defects on PET/SPECT imaging. This finding demonstrates the potential of radiomic filtering to provide hypothesis-generating data for future studies, and could eventually serve as a clinically useful tool to potentially help diagnose and screen for lung disease.

## Acknowledgments

The authors would like to thank Dr. John Kipritidis, Dr. Paul Keall, Dr. Shankar SIVA, Dr. Tokihiro Yamamoto, and Dr. Joseph Reinhardt for sharing the VAMPIRE dataset.

## Disclosures

## Data Availability

All imaging data used in this study is obtained from the publicly available VAMPIRE dataset [3]. The original data sets can be downloaded via request from the VAMPIRE website: https://image-x.sydney.edu.au/vampire-challenge/.